\documentclass[aps,prb,onecolumn,showpacs,showkeys,groupedaddress,floatfix,amsmath,amssymb]{revtex4}
\usepackage{graphicx}
\usepackage{dcolumn}
\usepackage{bm}

\begin{document}

\title{Microscopic model for multiple flux transitions in mesoscopic superconducting loops}

\author{D. Y. Vodolazov$^{1,2}$}
\email{vodolazov@ipm.sci-nnov.ru}
\author{F. M. Peeters$^2$}
 \affiliation{$^1$ Institute for Physics of Microstructures, Russian Academy of
Sciences, 603950, Nizhny Novgorod, GSP-105, Russia\\
$^2$ Departement Fysica, Universiteit Antwerpen, B-2020
Groenenborgerlaan 171, Antwerpen, Belgium }

\author{T. T. Hongisto}
\author{A. Yu. Arutyunov}
\affiliation{University of Jyv\"{a}skyl\"{a}, Department of
Physics, PB 35, 40014 Jyv\"{a}skyl\"{a}, Finland}


\begin{abstract}

A microscopic model is constructed which is able to describe
multiple magnetic flux transitions as observed in recent ultra-low
temperature tunnel experiments on an aluminum superconducting ring
with normal metal - insulator - superconductor junctions [Phys.
Rev. B \textbf{70}, 064514 (2004)]. The unusual multiple flux
quantum transitions are explained by the formation of metastable
states with large vorticity. Essential in our description is the
modification of the pairing potential and the superconducting
density of states by a sub-critical value of the persistent
current which modulates the measured tunnel current. We also
speculate on the importance of the injected non-equilibrium
quasiparticles on the stability of these metastable states.

\end{abstract}

\maketitle

When a superconducting loop is exposed to a perpendicular magnetic
field the energy $E_{L}$ of the $L$-th state is given by:
\begin{equation}
E_{L} \sim  \frac  {1}  {S}  \left( \frac {\Phi} {\phi _{0}} + L \right) ^{2},
\end{equation}
where $L=1/2\pi \oint \nabla \theta ds$ is the vorticity, $\theta$
is the coordinate-dependent phase of the superconducting order
parameter, and the integration is made along the contour of the
loop). S is the loop's circumference, $\Phi$ is the magnetic flux
through the area of the loop, and $\phi _{0}=h/2e$ is the
superconducting flux quantum. If the system can relax to its
ground state, then sweeping the magnetic field causes a periodic
variation of the kinetic properties with period $\Delta\Phi = \phi
_{0}$ corresponding to transitions $\Delta L = \pm 1$. The
persistent current in the loop is proportional to the derivative
of the energy $I \sim dE_{L} / d \Phi$, and shows the
characteristic sawtooth behavior with the same period. The
'switching' supercurrent density corresponding to transitions from
the state with vorticity $L$ to the nearest state $L\pm 1$ is
$j_{switch} \sim 1 / S$. This $\Delta\Phi  = \phi _{0}$ behavior
is commonly observed at temperatures close to the critical one
\cite{Little-Parks}.

Recently, low temperature experiments ($T \ll T_{c}$) were
performed on superconducting loops \cite {Pedersen, Vodolazov,
Arutyunov, Bourgeois} leading to jumps with vorticity changes of
$\Delta L>1$. A phenomenological explanation, based on a numerical
solution of the time-dependent Ginzburg-Landau equations, was
given in Ref. \cite{Vodolazov}. The basic idea of that theory was
as follows: at sufficiently low temperature and when sweeping the
magnetic field up the system can be 'frozen' in a metastable state
with vorticity $L_1$. When the circulating persistent current
reaches the corresponding critical density $j_{c} \sim 1 / \xi$
this metastable state relaxes to the low-energy level $L_2 \gg
L_1$. Thus, in loops with perimeter $S \gg \xi$ one may observe
vorticity changes $\Delta L \sim j_{c} / j_{switch} \sim S / \xi
\gg 1 $. One should notice that the explanation based on the
Ginzburg - Landau formalism is strictly valid only at temperatures
close to the critical one, and the extrapolation to the
low-temperature limit requires further justification. Therefore,
of particular interest are ultra-low temperature experiments
\cite{Arutyunov} where Al loops with circumference from a few $\mu
m$ to a few hundred $ \mu m$ were studied, and a well-defined
periodic structure with very large vorticity changes $\Delta L$ up
to $\sim$ 50 were observed. In the present paper we will construct
a microscopic model based on the Usadel equations \cite{Uzadel,
Belzig} to analyze the experiments of Ref.\cite {Arutyunov}.

Within the Usadel formalism, superconducting correlations between
electrons forming a Cooper pair are described by two complex
functions: the pairing angle $\theta$ and the phase $\chi$. In the
most general case, both functions depend on the coordinate $\bf
{r}$ and the energy $E$. The pairing angle and the phase are
linked through the coupled system of equations:
\begin{subequations}
\begin{eqnarray}
\frac{\hbar D}{2} \nabla^2 \theta+[i E-\frac{\hbar}{2D} {\bf
v}_s^2\cos \theta ]\sin \theta+\Delta \cos \theta=0, \quad
\\ 
\nabla({\bf v}_s \sin^2\theta)=0, \quad 
\end{eqnarray}
\end{subequations}
where the superfluid velocity is ${\bf v}_s=D(\nabla \chi
-(2e/\hbar){\bf A})$, $D$ is the diffusion coefficient and ${\bf
A}$ is the vector potential. The relation between the normal state
density of states at the Fermi level $N(0)$ and the
superconducting density of states is given by
$N(E)=N(0)Re[cos(\theta(E))]$. The pairing potential $\Delta ({\bf
r})$ should be defined self-consistently from the integral
equation:
\begin{equation}
\Delta({\bf r})=N(0)V_{eff}\int_0^{\hbar \omega_D}
\tanh(E/2k_BT)Im(\sin \theta) dE,
\end{equation}
where $V_{eff}$ is the pairing interaction strength and $\omega_D$
is the Debye frequency. These equations are valid for diffusive
superconductors with a mean free path $\ell$ much less than the
'pure limit' coherence length $\xi_0=0.18 \hbar v_F/ \Delta_0$
(where $v_F$ is the Fermi velocity of the electrons on the Fermi
surface and $\Delta_0$ is the gap at zero temperature, zero
magnetic field and in the absence of transport current).

For loops consisting of a sufficiently narrow wire $\surd {\sigma}
\leq \xi$ one may neglect the variation of the superconducting
properties in the transverse direction, where $\sigma$ is the
cross section of the wire. As a first approximation one may also
neglect the influence of the tunnel contact on the properties of
the superconducting loop \cite{Arutyunov}. Within these
approximations we obtain from Eqs. (2) and (3) a single algebraic
equation:
\begin{equation}
E+i\Gamma\cos \theta=i\Delta \frac{\cos \theta}{\sin \theta}
\end{equation}
where $\Gamma=\hbar \langle {\bf v}_s^2\rangle / 2D$ is the
depairing energy averaged across $\sigma$. One may neglect the
self-induced magnetic field (for the largest loop with $S=$100
$\mu$m it leads to an error less than $1\%$), and consider the
vector potential $\bf A$ as entirely due to the applied field
${\bf A}=(Hy/2,-Hx/2,0)$. For arbitrary vorticity $L$ in a
rectangular loop the depairing energy $\Gamma$ is:
\begin{equation}
2\Gamma/\Delta_0=\left(\frac{\pi L}{2}
\right)^2\frac{1}{(w-2d)w}-\frac{\pi
LH}{4}+\frac{H^2}{16}(w^2-2wd+4d^2/3)
\end{equation}
where the magnetic field $H$ is measured in units of
$H_{c2}=\Phi_0/2\pi\xi(0)^2$, $w$ is the side of the loop and $d$
is the width of the wire the loop is made of and both are measured
in units of $\xi(0)=\sqrt{\hbar \Delta_{0} / D}$.
\begin{figure}[h]
\includegraphics[width=\textwidth]{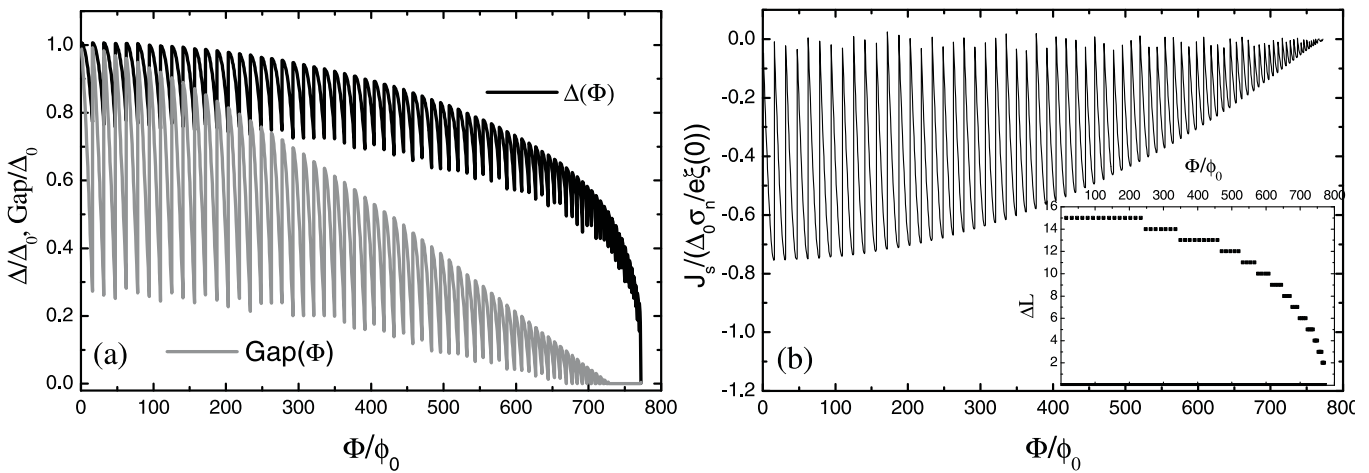} \caption {Dependence of
the gap (a), pairing potential $\Delta$ (a), current density (b)
and jump in vorticity $\Delta L$ (inset in Fig. 1(b)) on the
applied magnetic field in a superconducting loop with parameters:
T$_c$=1.2 K, $\xi$(0)=150 nm, w=5 $\mu m$ , d=120 nm, T=106 mK.}
\end{figure}

It was shown \cite{Vodolazov2} that for quasi-one-dimensional
superconducting loops of large circumference ($S \gg \xi$)
metastable states with fixed vorticity $L_n$ become unstable if
the persistent current density is equal to the depairing current
density $j_{c}$. In the present calculation we assume that when
the current density in the loop (being a function of magnetic
field $H$) reaches its maximum (at a fixed value of vorticity L)
the loop switches to a new quantum state. The new value of the
vorticity $L_m$ is then found from the condition that the pairing
potential $\Delta$ (minimum depairing energy $\Gamma$) for a given
magnetic field is maximal. In terms of a quasiclassical
description this transition criterion corresponds to the
supervelocity $v_s$ being equal to its critical value. In Fig. (1)
the calculated dependencies of the superconducting gap, pairing
potential $\Delta$ \cite{our2} and superconducting current density
are plotted as function of the applied magnetic field for a
superconducting loop with parameters similar to the ones used in
Ref. \cite{Arutyunov}. Of particular interest is the fact that for
small magnetic fields the period of oscillations is much larger
than $\phi_{0}$ and this period decreases in higher fields. Both
observations are qualitatively consistent with the recent
experiments \cite{Pedersen, Vodolazov, Arutyunov}. However,
simulations systematically give larger vorticity jumps than the
ones observed in the experiment although provide the same
dependence on the loop size (compare Eq. (6) and Fig. (7) in Ref.
\cite{Arutyunov}). Reasonable explanation of the discrepancy is
the presence of inevitable imperfections acting as weak links in
real samples. Even at the lowest temperatures the system exhibits
transitions from the 'frozen' metastable states at smaller values
of the magnetic field. In Ref. \cite{Vodolazov} it was shown that
an inhomogeneity in the loop can indeed strongly decrease the
values of the actually observed vorticity changes. Close to the
critical temperature thermal fluctuations disable the formation of
metastable states, which results in the 'conventional' $\phi_{0}$
periodicity ($\Delta L =\pm1$) \cite{Little-Parks, Arutyunov2}.


It is interesting to compare the results of the present
microscopic approach with calculations based on the
Ginzburg-Landau model \cite{Vodolazov,Vodolazov2}. For a
rectangular loop with side $w$ and thickness $d$ ($w \gg \xi \geq
d$) at $T/T_{c}\ll1$ the transition criterion for a state with
vorticity $L=0$ corresponds to the condition
$\Gamma_c/\Delta_0\sim 0.235$ (at $T \rightarrow 0$). Then the
size of the vorticity jump is:
\begin{equation}
\Delta L={\rm Nint} \left(
\sqrt{\frac{2\Gamma_c}{\Delta_0}}\frac{2w}{\pi \xi}\right)\simeq
{\rm Nint} \left(0.68\frac{2w}{\pi \xi} \right)
\end{equation}
where Nint(x) returns the integer value of the real variable x.
The solution of the Ginzburg-Landau equations \cite{Vodolazov2}
for the same transition gives:
\begin{equation}
\Delta L^{GL}={\rm Nint} \left(\frac{1}{\sqrt{3}}\frac{2w}{\pi
\xi}\right) \simeq {\rm Nint} \left( 0.58\frac{2w}{\pi \xi}
\right).
\end{equation}
The slight numerical discrepancy between Eqs. (6) and (7) is due
to the different functional dependence of $\Gamma _{c}$ in the
Ginzburg-Landau model and the present one. The quantitative
difference between Eqs. (6) and (7) is insignificant compared to
experimental inaccuracies \cite {Pedersen, Vodolazov, Arutyunov,
Bourgeois}.
\begin{figure}[h]
\includegraphics[width=\textwidth]{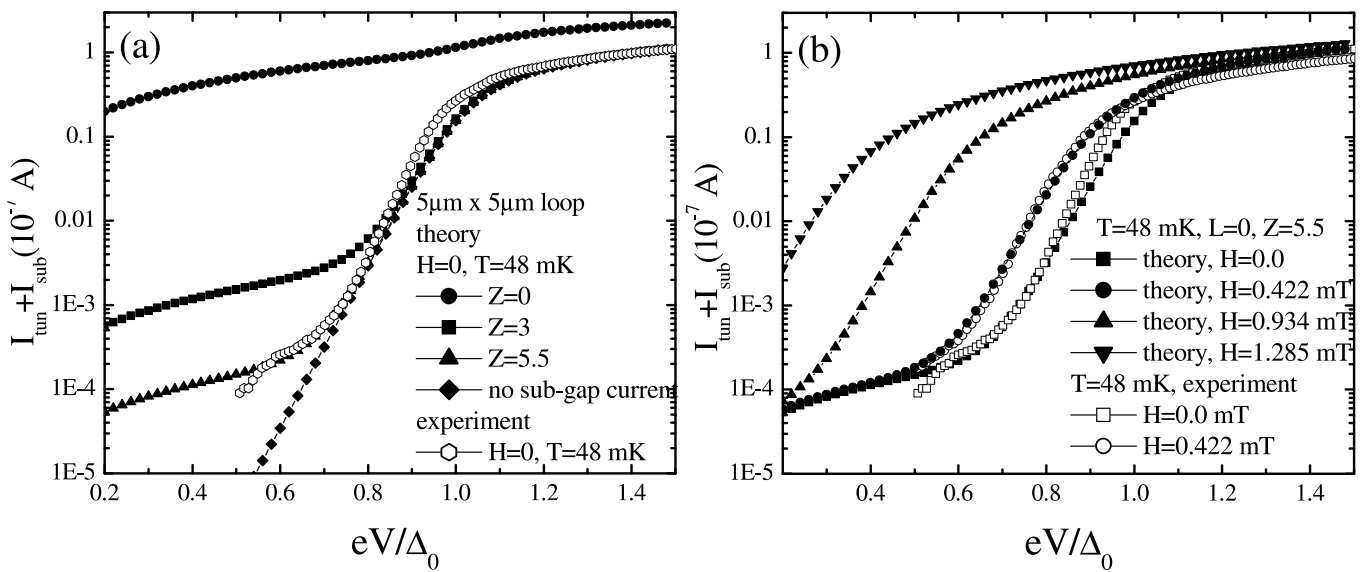} \caption{ Dependencies of the tunnel
current on the applied voltage: (a) at various strength of the
tunnel barrier, zero magnetic field and fixed temperature, (b) at
various magnetic fields, fixed temperature and tunnel barrier
strength.}
\end{figure}

So far we have considered persistent currents in an isolated
superconducting loop. Periodic modulation of these currents by an
external magnetic field can be measured using magnetization
\cite{Pedersen, Vodolazov} or calorimetric \cite{Bourgeois}
methods. In the ultra-low temperature experiments of Ref.
\cite{Arutyunov} an Al loop was used as the superconducting (S)
electrode of a  N-I-S junction being overlaped through a tunnel
barrier (I) with a normal-metal (N) contact. The tunnel current
$I_{tun}$ was measured at fixed bias voltage (or the voltage
across the barrier at a fixed tunnel current) as a function of the
applied perpendicular magnetic field. We argue that the observed
periodic variation of the tunnel current (voltage) with magnetic
field \cite{Arutyunov} is due to the modulation of the gap and the
pairing potential by the sub-critical persistent current
\cite{Anthore} in the loop-shaped superconducting electrode of the
N-I-S structure. To calculate the tunnel current we use the
conventional 'semiconductor' model \cite{Tinkham} which gives for
the tunnel current:
\begin{equation}
I_{tun}=\alpha|T|^2 \times
\int_{-\infty}^{\infty}\frac{N(E)}{N(0)}
[ f(E,T) - f(E+eV,T) ] dE
\end{equation}
where $f(E,T)$ is the Fermi distribution function,  $\alpha$ is
the constant of proportionality, and $T$ is the tunnelling matrix
element. It can be shown that $\alpha |T|^2 = 1 / eR_T$, where
$R_T$ is the tunnel resistance of the junction in the normal state
and $e$ is the electron charge \cite{Tinkham}. In our case of
strong currents, to obtain the density of states in the
superconductor $N(E)$ we solved Eqs. (2-5) self-consistently.

The described approach provides a qualitative description of the
main features observed in the experiment of Ref. \cite{Arutyunov}:
multi-flux periodicity of the tunnel current and the decrease of
the period of the tunnel current oscillation with increasing
magnetic field. However, quantitative comparison is far from being
perfect: the measured current of the voltage biased N-I-S junction
in the limit $eV<\Delta$ is systematically higher than the value
calculated using Eq. (8) (see Fig. 2(a)). We believe that to
obtain a better quantitative agreement one should consider
higher-order processes describing tunnelling of pairs of electrons
\cite{Blonder,Hekking} giving rise to the sub-gap current. To
account for this process we use a combination of the results
obtained in Ref. \cite{Blonder} for pure superconductors
(quasi-particle transport in the ballistic regime) and the simple
'semiconductor' model \cite{our1}. This leads to the
semi-quantitative result for the sub-gap current:
\begin{equation}
I_{sub}=\alpha|T|^2 \times
\int_{-\infty}^{\infty} A(\Delta,E)
[ f(E,T) - f(E+eV,T) ] dE
\end{equation}
where the probability of Andreev reflection $A(\Delta,E)$ is taken
from Table II of Ref. \cite{Blonder}:
\begin{equation}
A(\Delta,E)=\left \{ \begin{array}{ll}
 \displaystyle{\frac{\Delta^2}{E^2+(\Delta^2-E^2)(1+2Z^2)^2}}, & E<\Delta,\\
 \displaystyle{\frac{\Delta^2}{(E^2-\Delta^2)(\sqrt{E^2/(E^2-\Delta^2)}+2Z^2+1)^2}}, & E>\Delta,
\end{array} \right.
\end{equation}
with $Z$ being the dimensionless barrier strength \cite{Blonder},
and $\Delta$ is found from solutions of Eqs. (2-5). The tunnel
current $I_{tun}$ and the sub-gap current $I_{sub}$ have different
functional dependencies on $\Delta$, and hence their variation in
magnetic field is also different. While sweeping the magnetic
field the magnitude of the oscillations of the tunnel current is
much higher than the magnitude of oscillations of the sub-gap
current. Strictly speaking, this over-simplified introduction of a
sub-gap current in our model is not applicable for dirty limit
electrodes studied in Ref. 4. However, it demonstrates that at
energies $\Delta << eV$  an account for the sub-gap current
contribution gives qualitatively better agreement with experiment:
Fig. 2 and Fig. 3. The described approach is not able to give
quantitative exact values for the sub-gap current as other
mechanisms (e.g. leakage current) might contribute to the total
current measured in N-I-S structures. It is also known that the
magnitude of the sub-gap current strongly depends on interference
effects within the locus of the N-I-S junction, and hence is
geometry-dependent \cite{Hekking}.

\begin{figure}[h]
\includegraphics[width=\textwidth]{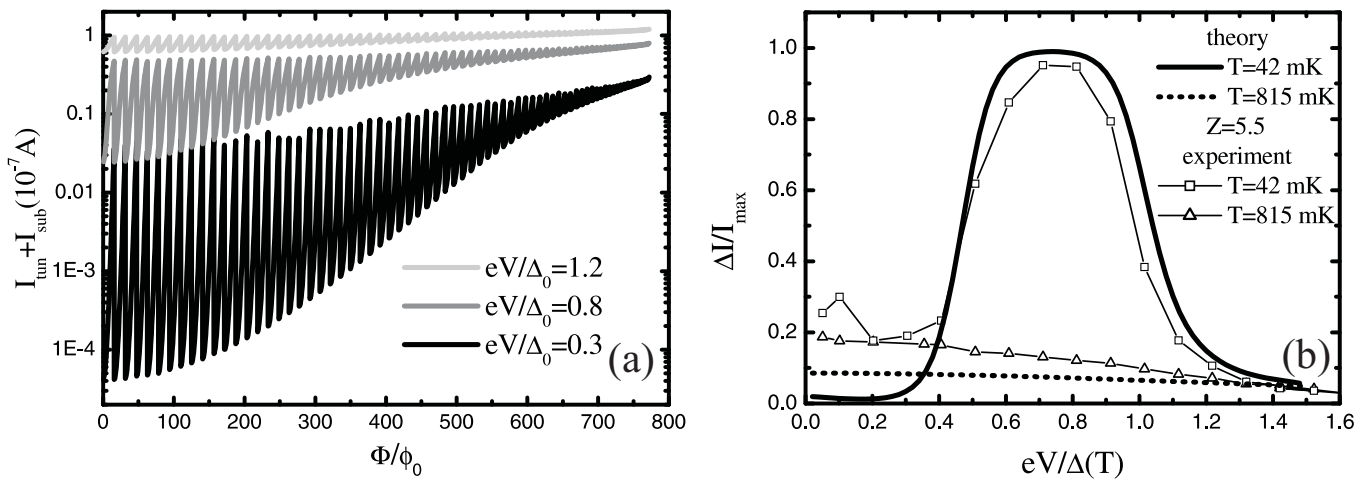} \caption{Magnetic field dependence of
the total current through a N-I-S junction for different values of
the bias voltage (a) and normalized magnitudes of the low field
($B <$ 5 mT) current oscillations $\Delta I/I_{max}$ as function
of the normalized bias $eV/\Delta(T)$ (b). Parameters of the loop
are the same as in Fig. 1.}
\end{figure}
In spite of the obvious simplifications when describing the
sub-gap current, Eqs. (8-10) combined with calculations of the
density of states (Eqs. (2-5)), is able to give good qualitative
and reasonable quantitative agreement with experiment
\cite{Arutyunov}. As follows from Fig. 2, and in full agreement
with the experimental findings \cite{Arutyunov}, the absolute
value of the magnitude of the current oscillations as function of
the magnetic field $I(B,V=const)$ (Fig. 3(a)) and their normalized
magnitude $\Delta I/I_{max}(B,V=const)$ (Fig. 3(b)) depend on the
bias voltage $V$. The only 'tuning' parameter used in fitting the
calculations with experiment is the coefficient $Z$ (see Fig.
2(a)).

At high temperatures the current across a tunnel junction at $V
\ll \Delta$ is mainly determined by the tunnel component Eq. (8)
due to temperature smearing of the Fermi distribution function.
While at very low temperatures the contribution of the tunnel
current is negligible in the same limit, and the finite measured
current is practically equal to the sub-gap term of Eqs. (9,10).
For all temperatures and $V>\Delta$ the total current is
determined by the tunnel component. In view of the relatively weak
dependence of the sub-gap current (see Eqs. (9,10)) on magnetic
field, this explains qualitatively the existence of a maximum in
the dependence of $\Delta I/I_{max}$ versus voltage at low
temperatures, and its absence at higher temperatures (Fig. 3(b)).

In order to improve our model one should include the injection of
non-equilibrium quasiparticles from the normal electrode.
Particularly at energies $eV \sim \Delta$ this effect results in a
cooling of the normal-metal contact and a heating of the
superconducting contact \cite{Nahum, Arutyunov3}. The inevitable
consequence is a spatially-inhomogeneous modification of the
superconducting gap within the locus of the N-I-S junction.
Manifestation of such an effect has been observed in S-I-S'-I-S
structures at temperatures comparable to the critical temperature
of the S' electrode \cite{Blamire, Heslinga}. At $T\rightarrow 0$
there are very few equilibrium (thermally activated) excitations,
and one might expect that even a small amount of extra
quasiparticles injected from the normal electrode may affect the
pairing potential $\Delta$ and, hence, influence the stability of
the 'frozen' metastable states. We speculate that the mentioned
non-equilibrium effect might explain also the variation of the
period $\Delta B_{I}$ of the current oscillation on the bias
voltage (Fig. 6 and Fig. 8(c) in Ref. \cite{Arutyunov}). The
larger the applied voltage, the higher the quasiparticle
injection, and, hence, the stronger the deviation of the
distribution function in the superconducting electrode from its
equilibrium value. Therefore a metastable state with a given
vorticity becomes unstable at smaller values of the depairing
energy $\Gamma$, or, in other words, at smaller values of the
applied magnetic field.

In conclusion, using a microscopic approach we analyzed the
ultra-low temperature behavior of a mesoscopic-size
superconducting ring in the presence of a magnetic field. The
model was used to interpret  recent experiments \cite{Arutyunov}
on N-I-S junctions with a loop-shaped superconducting electrode.
The central result of the present work is the demonstration that
the tunnel current oscillates in a magnetic field with a period,
which scales with the loop circumference. For large loops flux
changes are much larger than a single flux quantum. We found
agreement with experiment. One should include the finite sub-gap
current in order to improve the quantitative agreement between the
present model and the measured total current across the junction.
Using simple assumptions, we were able to obtain a reasonable
quantitative agreement with the experimental results of Ref.
\cite{Arutyunov}.

\begin{acknowledgements}

The work was supported by the EU Commission FP6 NMP-3 project
505587-1 "SFINX", the Russian Foundation for Basic Research (Grant
04-02-17397-A), Academy of Finland under the Finnish Center of
Excellence Program  2000-2005 No. 44875, Nuclear and Condensed
Matter Program at JYFL, the Belgian Science Policy, the Flemish
Science Foundation (FWO-Vl) and the ESF-network: AQDJJ. One of the
authors (D.Y.V.) is supported by INTAS Grant N 04-83-3139.

\end{acknowledgements}


\begin{thebibliography}{20}
\bibitem{our2} In general the pairing potential $\Delta$
does not coincide with the value of the superconducting gap.
$\Delta$ determines the superconducting correlations, which do not
necessary lead to the appearance of an energy gap in the spectrum
of excitations. The latter is defined as a threshold energy at
which the density of states turns finite.
\bibitem{our1} The model of Ref. \cite{Blonder} was based on the
Bogoliubov - de Gennes equations for ballistic transport through a
tunnel junction. The results coincide with the simple
'semiconductor' model in the absence of Andreev reflection and for
low-transparent tunnel junctions. Therefore we believe  that the
described approach accounts for the sub-gap current at least
semi-quantitatively.
\end{thebibliography}
\end{document}